\documentclass[preprint,showpacs,showkeys]{revtex4}
\usepackage{amsmath,amsfonts,amssymb}

\begin{document}

\title{On the boundary-value problems and the validity of the Post
constraint in modern electromagnetism}
\author{Yuri N. Obukhov}
\email{yo@thp.uni-koeln.de}
\affiliation{Institute for Theoretical Physics, University of Cologne, 
50923 K\"oln, Germany\,}
\altaffiliation[Also at: ]{Department of Theoretical Physics, Moscow 
State University, 117234 Moscow, Russia}
\author{Friedrich W. Hehl}
\email{hehl@thp.uni-koeln.de}
\affiliation{Institute for Theoretical Physics, University of Cologne, 
50923 K\"oln, Germany}
\altaffiliation[Also at: ]{Department of Physics \& Astronomy, 
University of Missouri-Columbia, Columbia, MO 65211, USA}

\begin{abstract}
We recall that the theory of electromagnetism consists of three building
blocks: (a) the inhomogeneous Maxwell equations for the electric and 
magnetic excitations $({\bf D}, {\bf H})$ (which reflects charge conservation),
(b) the homogeneous Maxwell equations for the electric and magnetic field
strengths $({\bf E},{\bf B})$ (which reflects flux conservation), and 
(c) the constitutive relation between $({\bf D}, {\bf H})$ and $({\bf E},
{\bf B})$. In the recent paper \cite{Lakhtakia1}, Lakhtakia proposed to change 
the standard boundary conditions in electrodynamics in order to exclude 
certain constitutive parameters. We show that this is inadmissible both 
from the macroscopic and the microscopic points of view. 
\end{abstract}

\maketitle

\section{Introduction}

Let us consider magnetoelectric matter with the constitutive relation
\begin{eqnarray}
{\bf D} &=& \varepsilon\varepsilon_0\,{\bf E} + \alpha\,{\bf B},\label{DE}\\
{\bf H} &=& {\frac 1{\mu\mu_0}}\,{\bf B} - \alpha\,{\bf E}.\label{HB}
\end{eqnarray}
Here $\varepsilon_0$ and $\mu_0$ are the electric and magnetic constants of
the vacuum, $\varepsilon$ and $\mu$ are the permittivity and permeability of
the matter, and $\alpha$ is the axion (Tellegen) parameter. 

In a recent discussion of the electrodynamics in magnetoelectric 
media of such a special type, Lakhtakia \cite{Lakhtakia1} proposed to replace 
the standard boundary (jump) conditions on an interface between two media 
characterized by different values of the constitutive parameters $\varepsilon,
\mu, \alpha$ by new boundary conditions which do not contain the axion 
parameter $\alpha$.

The motivation for such a replacement is as follows. When the medium is completely
homogeneous, it is easy to verify that Maxwell's differential equations do not
contain the parameter $\alpha$ when (\ref{DE})-(\ref{HB}) is substituted into
these equations. However, for the inhomogeneous situation, when there are two 
spatial domains with different values of $\alpha$,  for example, one should
implement the boundary conditions on the surface that separates the two domains. 
The use of the standard boundary conditions allows for the influence of the 
nontrivial axion field $\alpha$ on the physical processes (in particular, on 
waves propagating through the interface). Lakhtakia \cite{Lakhtakia1} modified
the jump conditions such that this effect of $\alpha$ is removed. 

The authors of a comment \cite{Sihvola} gave arguments against such an 
arbitrary modification of the boundary conditions. However, Lakhtakia 
\cite{Lakhtakia2} disagreed, referring to the difference between the 
microscopic and macroscopic approaches to electrodynamics. Here we 
show that the claim of Lakhtakia is incorrect, both macroscopically and
microscopically.

\section{Microscopic and macroscopic electrodynamics}

We start from the microscopic Maxwell equations (see eqs. (1)-(4) of 
\cite{Lakhtakia1}) that, after the spatial averaging, yield the equations 
of macroscopic electrodynamics:
\begin{eqnarray}
\varepsilon_0\,{\bf \nabla}\cdot{\bf E} = \rho,&\qquad& \mu_0^{-1}\,
{\bf \nabla}\times{\bf B} - \varepsilon_0\,\dot{\bf E} = {\bf J},\label{Mi}\\
{\bf \nabla}\cdot{\bf B} = 0,&\qquad& 
{\bf \nabla}\times{\bf E} +\dot{\bf B} = 0.\label{Mh}
\end{eqnarray}
Here we agree completely with Lakhtakia, namely that all four dynamical equations 
contain only two fields, ${\rm E}$ and ${\bf B}$, and that all four equations 
hold in vacuum as well as in matter. There is however, an important difference:
whereas in vacuum the electric charge and current densities $\rho$ and ${\bf J}$
are defined by the {\it free} charges only, in matter the sources $\rho$ and 
${\bf J}$ are sums of both {\it free} and {\it bound} charges:
\begin{equation}
\rho = \rho^{\rm f} + \rho^{\rm b},\qquad {\bf J} = {\bf J}^{\rm f} 
+ {\bf J}^{\rm b}.\label{rj}
\end{equation}
The bound sources are related to the polarization and magnetization which 
provide an averaged description of the physical response of the microscopic
constituents of matter exposed to the action of the electric and magnetic 
fields:
\begin{equation}
\rho^{\rm b} = - {\bf \nabla}\cdot{\bf P},\qquad 
{\bf J}^{\rm b} = {\bf \nabla}\times{\bf M} +\dot{\bf P}.\label{PM}
\end{equation}
Substituting (\ref{rj}) into (\ref{Mi}), and making use of (\ref{PM}), we 
recast the inhomogeneous Maxwell equations into 
\begin{equation}
{\bf \nabla}\cdot{\bf D} = \rho^{\rm f},\qquad {\bf \nabla}\times{\bf H} 
- \dot{\bf D} = {\bf J}^{\rm f},\label{Mim}
\end{equation}
with the electric and magnetic excitation fields defined by
\begin{equation}
{\bf D} = \varepsilon_0{\bf E} + {\bf P},\qquad {\bf H} = \mu_0^{-1}{\bf B}
- {\bf M}.\label{DH} 
\end{equation}

\section{Boundary conditions}

Let us consider the case when space is divided into two domains by a
plane boundary surface $S$. When matter in each of the domains is homogeneous,
it is convenient technically to study the electrodynamical processes (waves,
in particular) in the two half-spaces separately. However, if we want the 
physical picture in the whole space, we will need boundary (jump) conditions 
for the electric and magnetic fields across the surface $S$. 

We derive here the boundary conditions in a slightly different way than it is
done in textbooks. Namely, we will do it directly for the fundamental fields 
${\bf E}$ and ${\bf B}$ and not for the excitations. Since each of the fields
is a vector, we expect three boundary conditions for each field (one for the
normal projection and two for the tangential components). Indeed, with the 
help of the usual technique by integrating the Maxwell equations 
(\ref{Mi})-(\ref{Mh}) in a thin region in a small vicinity of the surface $S$,
we derive the conditions for the fields at the boundary:
\begin{eqnarray}
{\bf E}_n^{(2)} - {\bf E}_n^{(1)} = \varepsilon_0^{-1}\rho_s,\qquad 
{\bf E}_\tau^{(2)} - {\bf E}_\tau^{(1)} = 0,\label{deltaE}\\
{\bf B}_\tau^{(2)} - {\bf B}_\tau^{(1)} = \mu_0{\bf J}_s,\qquad 
{\bf B}_n^{(2)} - {\bf B}_n^{(1)} = 0.\label{deltaB}
\end{eqnarray}
The notation is obvious: the subscripts ${}_n$ and ${}_\tau$ denote the normal 
and tangential projections, whereas the superscripts ${}^{(1)}$ and ${}^{(2)}$ 
label the half-space domains. 

The crucial feature of the boundary conditions across $S$ is the presence of
the {\it surface charge and current} densities $\rho_s$ and ${\bf J}_s$ in 
(\ref{deltaE}) and (\ref{deltaB}). These surface densities have the form
\begin{equation}
\rho_s = \rho_s^{\rm f} + \rho_s^{\rm b},\qquad {\bf J}_s = {\bf J}_s^{\rm f}
+ {\bf J}_s^{\rm b},\label{rjS} 
\end{equation}
thus including on an equal
footing both the surface density of the {\it free} sources $\rho_s^{\rm f}$ 
and ${\bf J}_s^{\rm f}$ and the surface density of the {\it bound} sources 
$\rho_s^{\rm b}$ and ${\bf J}_s^{\rm b}$. The physical origins of the two
types of the surface charges and currents on $S$ are somewhat different. 
Whereas $\rho_s^{\rm f}$ and ${\bf J}_s^{\rm f}$ describe the possible 
presence of the free sources right at the boundary (prepared under the 
conditions of an experiment, for example), the surface sources $\rho_s^{\rm b}$ 
and ${\bf J}_s^{\rm b}$ arise from the fact that matter (from the microscopic
point of view) has different electromagnetic properties in the two half-spaces. 
As a result, the polarization and magnetization, while being continuous in
each separate homogeneous domain, are {\it discontinuous} across the boundary $S$.

We can easily derive the relation between the bound surface sources on $S$ 
and the jump of the polarization and magnetization across $S$. Integrating
the equations (\ref{PM}) in a infinitely thin neighbourhood including $S$
(in the same way like for the Maxwell equations), we derive
\begin{equation}
{\bf P}_n^{(2)} - {\bf P}_n^{(1)} = - \rho_s^{\rm b},\qquad 
{\bf M}_\tau^{(2)} - {\bf M}_\tau^{(1)} = {\bf J}_s^{\rm b}.\label{deltaPM}
\end{equation}
Substituting (\ref{rjS}) and (\ref{deltaPM}) into (\ref{deltaE})-(\ref{deltaB}),
and taking into account (\ref{DH}), we recast the boundary conditions in a
more familiar form:
\begin{equation}
{\bf D}_n^{(2)} - {\bf D}_n^{(1)} = \rho_s^{\rm f},\qquad 
{\bf H}_\tau^{(2)} - {\bf H}_\tau^{(1)} = {\bf J}_s^{\rm f}.\label{deltaDH}
\end{equation}

\section{Axion (Tellegen) magnetoelectric medium and the Post constraint}

Up to this point, we were quite general. Now, let us specialize to the case of 
the magnetoelectric matter with the constitutive relation (\ref{DE})-(\ref{HB}).
The permittivity, permeability, and the axion (Tellegen) field in the two 
half-space domains are $\varepsilon_1,\mu_1,\alpha_1$ and $\varepsilon_2,\mu_2,
\alpha_2$, respectively. Using (\ref{DE})-(\ref{HB}) in (\ref{deltaDH}) and 
(\ref{deltaE})-(\ref{deltaB}), we find the boundary conditions 
\begin{eqnarray}
\varepsilon_0\left(\varepsilon_2{\bf E}_n^{(2)}- \varepsilon_1{\bf E}_n^{(1)}
\right) + (\alpha_2 - \alpha_1){\bf B}_n^{(1)} &=& \rho_s^{\rm f},\label{a1}\\
\mu_0^{-1}\left(\mu_2^{-1}{\bf B}_\tau^{(2)}- \mu_1^{-1}{\bf B}_\tau^{(1)}\right)
- (\alpha_2 - \alpha_1){\bf E}_\tau^{(1)} &=& {\bf J}_s^{\rm f}.\label{a2}
\end{eqnarray}
As we see, the axion field shows up explicitly. The analysis of the wave 
propagation in such a magnetoelectric medium reveals the influence of the axion 
field on the reflected and transmitted wave \cite{axion1,Sihvola} (see
also \cite{lindell,jancewicz}). One can use this effect to measure the axion 
(Tellegen) field \cite{axion2}. 

Lakhtakia's proposal to change the standard boundary conditions in order to
eliminate the contribution of axion (thus proving the so called Post 
constraint \cite{axion1}) is physically unsubstantiated. One cannot 
treat the boundary conditions as a kind of supplementary conditions that one 
can choose arbitrarily (like the the fixing of the gauge, for example). The 
boundary conditions (\ref{deltaE})-(\ref{deltaB}) and (\ref{deltaDH}) {\it are 
the Maxwell equations} written in a different (integral and not differential) 
form for a specific region of space (i.e., for the infinitely thin neighbourhood 
of the boundary surface $S$). By changing a boundary condition, Lakhtakia 
actually {\it changes} the physical laws, namely, the Maxwell equations at the 
interface between the two media. 

An alternative (equivalent) explanation of the presence of the axionic terms 
in (\ref{a1})-(\ref{a2}) is as follows. Let us look at the right-hand sides 
of these equations. The {\it free surface density} sources $\rho_s^{\rm f}$ 
and ${\bf J}_s^{\rm f}$ arise because, originally, in the Maxwell equations 
(\ref{Mi}) there is a $\delta$-function distribution of the free charge and 
current densities of the type $\rho^{\rm f} \cong \rho_s^{\rm f}\,\delta
({\bf x} - {\bf x}(S))$ (and similarly for the free current density). The same
applies also to the {\it bound surface density} sources. Indeed, the axion 
field, considered on the whole space, is a step function 
\begin{equation}
\alpha(x) = \begin{cases}{\alpha_1,\quad{\rm for}\ x\in{\rm 1st\ domain}}\\ 
{\alpha_2,\quad{\rm for}\ x\in{\rm 2nd\ domain}}\end{cases}
\end{equation}
Using this function in (\ref{PM}), we find a $\delta$-function contribution
from the magnetoelectric piece in the polarization and magnetization, namely,
$\rho^{\rm b} \cong (\alpha_1 - \alpha_2)B_n\,\delta ({\bf x} - {\bf x}(S))$
(and a similar expression for the bound current density). 

When these two delta-functions (one for the free and another for the bound 
sources) are substituted into the Maxwell equations (\ref{Mi}), integration
in the infinitesimally thin region around the boundary $S$ yields the two
contributions to the right-hand side of (\ref{deltaE})-(\ref{deltaB}). The
free source delta-function is responsible for the surface density terms
$\rho_s^{\rm f}$ and ${\bf J}_s^{\rm f}$, whereas the bound source 
delta-function gives rise exactly to the axion terms in (\ref{a1})-(\ref{a2}).
A similar argument was used in the previous comment \cite{Sihvola}. In his
response \cite{Lakhtakia2}, Lakhtakia claimed that a microscopic approach
and a ``homogenization" of the fields might support his proposal. However,
here we have analysed the problem starting from a microscopic viewpoint. It 
is unclear how any kind of ``homogenization" can eliminate a delta-function 
at the boundary between the two domains filled with {\it different} matter. 

If we take any point at the boundary $S$ and perform averaging and
``homogenization" in an arbitrarily small neighourhood of this point, we will
necessarily find two portions of space to the left and to the right of $S$,
in which the electric and magnetic properties are homogeneous within the
respective portions of the neighbourhood, but are {\it not homogeneous} and 
even not continuous across $S$. There just cannot be any ``homogenization"
across the boundary since $S$ divides the two materials with essentially
different physical properties. For example, we can have vacuum in the 
first half-space and a magnetoelectric medium in the second half-space. 
The vacuum is not polarized and magnetized. In contrast, the magnetoelectric 
medium becomes electrically polarized in a magnetic field and/or becomes 
magnetized in electric field, with the parameter $\alpha$ determining 
such polarization and magnetization. As a result, ${\bf P}$ and ${\bf M}$
are both {\it discontinuous} across $S$, and ``homogenization" cannot
change this fact.

\section{Conclusion}

Classical macroscopic electrodynamics (which can be consistently derived
with the help of the spatial averaging from the microscopic electrodynamics)
consists of three building blocks \cite{book}: (a) the inhomogeneous Maxwell 
equations (\ref{Mi}) (which reflects charge conservation), (b) the homogeneous 
Maxwell equations (\ref{Mh}) (which reflects flux conservation), and (c) 
the constitutive relation between the electromagnetic field excitations
$({\bf D}, {\bf H})$ and the electromagnetic field strength $({\bf E}, 
{\bf B})$. The latter encodes the response of the medium to the action
of the electric and magnetic fields in terms of the polarization and 
magnetization fields that are related to the bound charge and current
source densities. Although some constitutive parameter (like the
axion field $\alpha$) may drop out of the differential equations (\ref{Mi}),
it still enters the constitutive law, reflecting the state of polarization
of matter induced by the magnetic field and/or the state of magnetization
induced by the electric field. 

A modification of the boundary (jump) conditions across the surface $S$
between the two different media, proposed in \cite{Lakhtakia1}, is 
physically inadmissible because such a change of the boundary conditions
amounts to a change of the Maxwell equations. Moreover, the averaging
and ``homogenization" arguments cannot eliminate the discontinuous 
behavior of the polarization and magnetization across the boundary $S$
which is manifest in the delta-function like contributions both to the sources 
of the {\it free charge} and the {\it bound charge}. The Post constraint is
unphysical and invalid.

{\bf Acknowledgments} This work was supported by the Deutsche 
Forschungsgemeinschaft (Bonn) with the grant HE 528/21-1.

\end{document}